August 1994

HUB-IEP-94/12

KL-TH 19/94

# Spectrum of the Dirac Operator and Multigrid Algorithm with Dynamical Staggered Fermions *

Thomas Kalkreuter **

*Fachbereich Physik*
*Universität Kaiserslautern*
*Erwin-Schrödinger-Straße*
*D-67663 Kaiserslautern, Germany*
and ‡
*Fachbereich Physik (Computational Physics)*
*Humboldt-Universität*
*Invalidenstraße 110*
*D-10099 Berlin, Germany*

**Abstract**

Complete spectra of the staggered Dirac operator $\not{D}$ are determined in quenched four-dimensional $SU(2)$ gauge fields, and also in the presence of dynamical fermions. Periodic as well as antiperiodic boundary conditions are used. An attempt is made to relate the performance of multigrid (MG) and conjugate gradient (CG) algorithms for propagators with the distribution of the eigenvalues of $\not{D}$. The convergence of the CG algorithm is determined only by the condition number $\kappa$ and by the lattice size. Since $\kappa$'s do not vary significantly when quarks become dynamic, CG convergence in unquenched fields can be predicted from quenched simulations. On the other hand, MG convergence is not affected by $\kappa$ but depends on the spectrum in a more subtle way.

*Work supported by Deutsche Forschungsgemeinschaft.

**Electronic address: kalkreut@linde.physik.hu-berlin.de

‡Address after May 1, 1994

# 1 Introduction

Big efforts have been undertaken to find efficient multigrid (MG) methods for the computation of propagators in background gauge fields [1, 2, 3, 4, 5, 6, 7, 8, 9, 10]. The goal is to find an improved method for the solution of discretized Dirac equations. Such an improvement would accelerate numerical simulations of theories involving dynamical fermions considerably when one uses the Hybrid Monte Carlo algorithm [11]. [1)]

Although ultimately one wants to simulate theories with dynamical fermions, all the works on MG methods mentioned above focussed only on quenched gauge fields. However, it is reasonable to expect that MG methods have a chance to perform better when one considers "real" gauge fields which are generated in the presence of dynamical fermions. On the other hand one will not expect any big difference for the behavior of the conjugate gradient (CG) algorithm. The reasons for these two statements are as follows. The inclusion of the fermionic determinant in the Monte Carlo process will tend to decrease the number of (approximate) zero modes. This is so because configurations with less low-lying modes are more probable. MG methods intend to take care of the low-lying modes (which are responsible for critical slowing down (CSD)) on coarser grids, and the task of dealing with a reduced number of low-lying modes should be easier. Concerning the CG algorithm, its (asymptotic) convergence properties are determined by the condition number. [2)] Since condition numbers of the (negative squared) massless Dirac operator are not influenced dramatically by the presence of dynamical quarks, one does not expect a significant consequence for the convergence behavior of CG.

Previously other works have been done which were concerned with the role of low-lying modes in MG methods. The Israel group made a visual study of approximate zero modes in the quenched Schwinger model in the framework of their multigrid algortihm [2]. The present author showed that there exists an "idealized" MG algorithm which is able to eliminate CSD (or at least to reduce CSD strongly) both for bosonic propagators [6] and for propagators of staggered fermions [9] in quenched four-dimensional $SU(2)$ gauge fields. A prerequisite for the success of the idealized MG method was the preservation of criticality of the Dirac operator under coarsening. In the present study we focus on the consequences of dynamical fermions for the performance of a simpler MG algorithm. This simpler algorithm proved to work in the quenched case, but it is unable to outperform CG [8]. Nevertheless, it is worth to see how it performs when dynamical fermions are present.

This paper is organized as follows. In Sec. 2 we recall the deterministic MG method. The spectrum of the staggered Dirac operator is determined by means of a Lanczos procedure. This procedure is recalled and numerical results are presented in Sec. 3. Computations of propagators by CG and by MG are reported in Sec. 4, and we end with some conclusions.

[1)] There is a recent alternative proposal by Lüscher [12] where one does not have to invert Dirac operators directly.

[2)] For a positive hermitean matrix the condition number equals the ratio of the largest to the smallest eigenvalue.

## 2 Multigrid method

We wish to solve the squared Dirac equation

$$(-\not{D}^2 + m^2)\,\chi = f \tag{1}$$

by MG, where $\not{D}$ is the gauge covariant staggered Dirac operator, $m$ is a bare mass parameter, and $f$ may be a pseudofermion field, for instance. We measure physical quantities in units of $a$, where $a$ denotes twice the lattice spacing of the staggered lattice. The reason for this is that free staggered fermions enjoy translational invariance only by shifts of $a$, not $a/2$ [13, 14].

The following MG notations will be used. The fundamental lattice is denoted by $\Lambda^0$. The first block lattice $\Lambda^1$ is obtained by coarsening with a factor of $L_b$. Thus $\Lambda^1$ has $L_b^d$ fewer sites than $\Lambda^0$ (in $d$ space-time dimensions). Restriction and interpolation operators $C$ and $\mathcal{A}$, respectively, are given by kernels $C(x,z)$ and $\mathcal{A}(z,x)$ with $z \in \Lambda^0$, $x \in \Lambda^1$. Note that $C(x,z)$ and $\mathcal{A}(z,x)$ are $N_c \times N_c$ matrices in a gauge theory with $N_c$ colors. Also, $C$ and $\mathcal{A}$ depend on the gauge field, although this is not indicated explicitly.

In a twogrid algorithm one performs a certain number, say one, of conventional relaxation sweeps on $\Lambda^0$, and one obtains an approximate solution $\tilde{\chi}$ of Eq. (1). The error $e = \chi - \tilde{\chi}$ is unknown, but the residual $r = f - (-\not{D}^2 + m^2)\,\tilde{\chi}$ is computable. Error and residual are connected by the residual equation $(-\not{D}^2 + m^2)\,e = r$. This equation is solved on $\Lambda^1$ where it reads [3)]

$$[C\,(-\not{D}^2 + m^2)\,\mathcal{A}]\,e_{\mathrm{block}} = C\,r \ . \tag{2}$$

Solving Eq. (2) is simpler than solving the original equation because there are fewer degrees of freedom on $\Lambda^1$. In the coarse grid correction step one replaces $\tilde{\chi}$ by $\tilde{\chi} + \mathcal{A}\,e_{\mathrm{block}}$. Then one performs again a relaxation sweep on $\Lambda^0$, etc. It is obvious how this twogrid algorithm can be generalized to a more-level MG algorithm.

We use a blocking procedure for staggered fermions which is consistent with the lattice symmetries of free fermions [14]. This forces us to choose $L_b = 3$. Even $L_b$ are not allowed. In four dimensions, coarsening by a factor of three reduces the number of points by 81. Therefore only a two-grid algorithm was implemented. This is sufficient to test the power of the MG method. The residual equation on the coarse grid was solved exactly by the CG algorithm.

The averaging kernel $C$ is chosen according to a ground-state projection definition. In the present work $C$ fulfills the gauge covariant eigenvalue equation(s) [17]

$$(-\Delta_{N,x} C^*)(z,x) = \lambda_0(x)\,C^*(z,x) \tag{3}$$

together with a normalization condition $CC^* = \mathbb{1}$, and a covariance condition $C(x,\hat{x}) \propto \mathbb{1}$ where $\hat{x}$ denotes the center of block $x$. In Eq. (3), $\lambda_0(x)$ is the lowest eigenvalue of $-\Delta_{N,x}$,

[3)] This is the Galerkin choice of the residual equation on the block lattice. It assumes that $e$ is smooth and can be obtained by smooth interpolation (via $\mathcal{A}$) of a suitable function $e_{\mathrm{block}}$ on $\Lambda^1$. The notion of smoothness in gauge fields is discussed in Ref. [8], and a more general discussion for disordered systems can be found in Ref. [15]; see also the recent work by Bäker [16].

and $-\Delta_{N,x}$ is the gauge covariant fermionic "two-link lattice Laplacian" – defined through $\not{D}^2 = \Delta + \sigma_{\mu\nu} F_{\mu\nu}$ – with "Neumann boundary conditions (b.c.)". $F_{\mu\nu}$ is the lattice definition of the field strength by means of plaquette terms. Neumann b.c. means that derivative terms in $\Delta$ are omitted where one site is in block $x$ and the other one is in a neighboring block. For more details we refer to Ref. [8].

The ground-state projection method is numerically implementable in four-dimensional non-abelian gauge fields [17], and since the method is gauge covariant, no gauge fixing in computations of propagators is required. Finally we cling to a variational method where the interpolation operator $\mathcal{A}$ equals the adjoint of the restriction operator $C$.

# 3 Spectrum of $-\not{D}^2$ in the presence of dynamical fermions

As explained in the introduction, naively one expects $\not{D}$ to have less approximate zero modes in the presence of dynamical fermions than in the quenched case. In order to study this conjecture we need firstly a Hybrid Monte Carlo program, and secondly a method to determine the low-lying spectrum of $\not{D}$. For the generation of four-dimensional $SU(2)$ gauge fields coupled to dynamical staggered fermions a FORTRAN program with vectorized CRAY code was used, which had been written by S. Meyer and B. Pendleton. They used this program when they studied the chiral transition with many fermion flavors in the $SU(2)$ Higgs model [18]. Meyer's and Pendleton's program was used with four flavors of staggered fermions. The spectrum of $\not{D}$ was determined by means of a Lanczos procedure to which we turn next.

## 3.1 Lanczos procedure

The Lanczos procedure is a technique that can be used to solve large, sparse, symmetric eigenproblems [19]. The method has been used in lattice field theory for a long time, see e. g. [20]. In course of the Lanczos iteration one generates for a given matrix $A$ a sequence of hermitean tridiagonal matrices $T^{(j)}$ by transformations with a matrix $Q$ whose columns are called "Lanczos vectors". These transformations have the property that the extremal eigenvalues of $T^{(j)}$ are progressively better estimates of the extremal eigenvalues of $A$. (In our case $A = -\not{D}^2$.) For details about the method and its convergence properties in exact arithmetic we refer to the literature [19].

In exact arithmetic, the Lanczos iteration should stop after at most $n$ steps when $A$ is an $n \times n$ matrix. In practice, however, there are severe problems [19, 21] caused by rounding errors and loss of orthogonality among the Lanczos vectors. As a consequence there appear so-called "spurious" eigenvalues [21] which are not eigenvalues of $A$. A clever way of identifying spurious eigenvalues and coping with their presence was proposed by Cullum and Willoughby [21]. In their algorithm one compares the eigenvalues of $T^{(j)}$ with the eigenvalues of a matrix $T_2$ which equals $T^{(j)}$ with the first row and first column deleted. If a simple eigenvalue of $T^{(j)}$ is also an eigenvalue of $T_2$, then this eigenvalue is spurious.

A problem which remains in the Cullum-Willoughby algorithm is that the correct multiplicities of the eigenvalues of $A$ do not emerge. This is so because a symmetric tridiagonal matrix does not have degenerate eigenvalues [23]. Nevertheless, in practice the matrices $T^{(j)}$ will have multiple eigenvalues which correspond to simple eigenvalues of $A$. They arise because the iteration essentially restarts itself when the numerical instabilities become too large. [4)]

In the present exploratory study the complete spectrum of $-\not{D}^2$ was determined. This was done in order to be sure about the correctness (i. e. to have no numerical uncertainties) in the distribution of low-lying modes. Of the several computational variants of the Lanczos procedure the most stable one as described in [19, Algorithm 9.2.1 and remark on p. 492] was implemented. Eigenvalues of the tridiagonal matrices $T^{(j)}$ were determined by means of the NAGLIB routine F02AVE [22].

## 3.2 Numerical Results in four-dimensional $SU(2)$ gauge fields

Let us first recall what one knows a priori about the multiplicities of the eigenvalues of $-\not{D}^2$. One knows that in $SU(2)$ gauge fields every eigenvalue is internally twofold degenerate[5)] [6]. Also, $-\not{D}^2$ couples only even lattice sites to even sites, and odd sites to odd sites. Therefore the spectrum $\mathcal{S}$ of $-\not{D}^2$ equals the union of the spectra $\mathcal{S}_{\text{even}}$ and $\mathcal{S}_{\text{odd}}$ of $-\not{D}^2$ restricted to the even and odd sublattices, respectively. One knows that $\mathcal{S}_{\text{even}} = \mathcal{S}_{\text{odd}}$ [9]. Thus, if there are no further degeneracies, then on a lattice $\Lambda$ of volume $|\Lambda|$ there must be $|\Lambda|/2$ different eigenvalues. Their sum must equal $\frac{1}{4}\text{Tr}\,(-\not{D}^2) = d|\Lambda|$. These statements are valid for periodic and for antiperiodic boundary conditions.

Concrete numerical investigations were done with the following parameters. The spectrum of $-\not{D}^2$ was investigated on $6^4$ and $12^4$ lattices. Two different kinds of boundary conditions (b.c.) were used. One with periodic b.c. in all directions for gauge and Fermi fields (denoted "per." in the following), and one with periodic b.c. for the gauge field in all directions and with antiperiodic b.c. on the Fermi field in time direction and periodic b.c. in spatial directions (denoted "antiper." in the following). The coupling $\beta = 4/g^2$ of the Wilson action for the $SU(2)$ gauge fields was varied between 1.8 and 5.0. Quark masses $m$ in the Hybrid Monte Carlo runs were chosen to be $m = 0.2$ and $m = 0.05$. These values were also used in Meyer's and Pendleton's work [18]; they quote $m = 0.1$ and $m = 0.025$ since they measured physical quantities in units of $a/2$.

In Cullum's and Willoughby's Lanczos procedure $j = |\Lambda|$ iterations were performed to compute $T^{(j)}$. (This is at least twice the number of "good" eigenvalues which can be expected.)

---

[4)]From the point of view of identifying correct eigenvalues, multiple eigenvalues of numerically computed $T^{(j)}$'s are welcome, because they are guaranteed to be not spurious.

[5)]I wish to thank U.-J. Wiese for pointing out to me that this degeneracy is due to a global charge conjugation symmetry which is special to $SU(2)$.

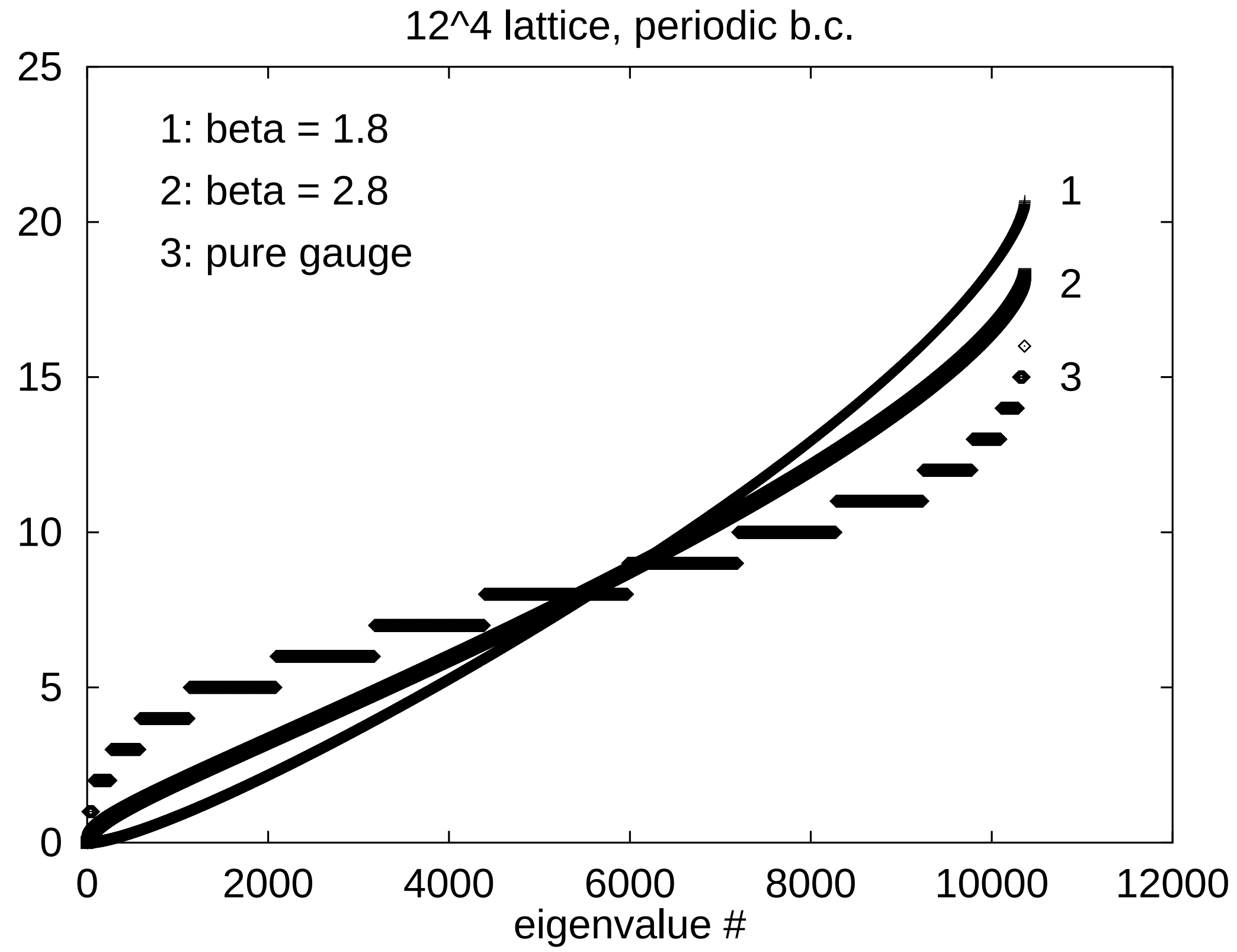

Figure 1: *Spectrum of* $-\not{D}^2$ *on* $12^4$ *lattices with periodic boundary conditions.*

The entries in the initial Lanczos vector were chosen randomly in $SU(2)$, and the vector was normalized in the 2-norm. Eigenvalues of $T^{(j)}$ were counted as being equal when they differed by less than $\varepsilon = 10^{-9}$. The same $\varepsilon$ was used to identify equal (simple) eigenvalues of $T^{(j)}$ and $T_2$. Proceeding in this way it turned out that in nontrivial gauge fields there seem to be no additional degeneracies to the ones explained above. Only for $\beta = 5.0$ it was impossible to identify $|\Lambda|/2$ eigenvalues whose sum equals $4|\Lambda|$. For $\beta = 1.8, \ldots, 2.8$ the method works perfectly. On $6^4$ ($12^4$) lattices we always found 648 (10368) eigenvalues whose sum came out as $5184 + \delta_6$ ($82944 + \delta_{12}$), with $\delta_6 < 8 \cdot 10^{-9}$ ($\delta_{12} < 1.7 \cdot 10^{-6}$) in REAL arithmetic on a CRAY Y-MP. Because of the randomness of the nonvanishing off-diagonal matrix elements of $-\not{D}^2$ this is very good evidence that the spectrum was determined exactly.

### 3.2.1 Complete Spectra

Spectra on $12^4$ lattices are shown in Figs. 1 and 2. We number the different eigenvalues $\lambda_k$ by $k = 0, 1, 2, \ldots$, with $\lambda_0 < \lambda_1 < \ldots$ The data shown for finite $\beta$ are results obtained with gauge fields generated by the Hybrid Monte Algorithm in the presence of dynamical fermions with a mass of $m = 0.2$. However, on the scale of the whole spectrum there is very little difference compared to $m = 0.05$ or to quenched gauge fields. The CPU time on a CRAY Y-MP is some 50 minutes for the implemented Lanczos procedure on $12^4$ lattices, and only 8 sec for $6^4$ lattices.

In Figs. 1 and 2 the data for $\beta = \infty$ (pure gauge) are not outcomes of numerical computations, but were taken from analytical results. In the free case the eigenvectors and eigenvalues of $-\not{D}^2$ are easily determined. For the eigenvalues $\lambda_k$ one finds on a $(2N)^d$ lattice with periodic b.c. [6)]

$$\lambda_k = 4\sum_{\mu=1}^{d} \sin^2\left(\frac{\pi k_\mu}{N}\right) \quad \text{with } k_\mu \in \{0, 1, \ldots, \left[\tfrac{N}{2}\right]\}. \tag{4}$$

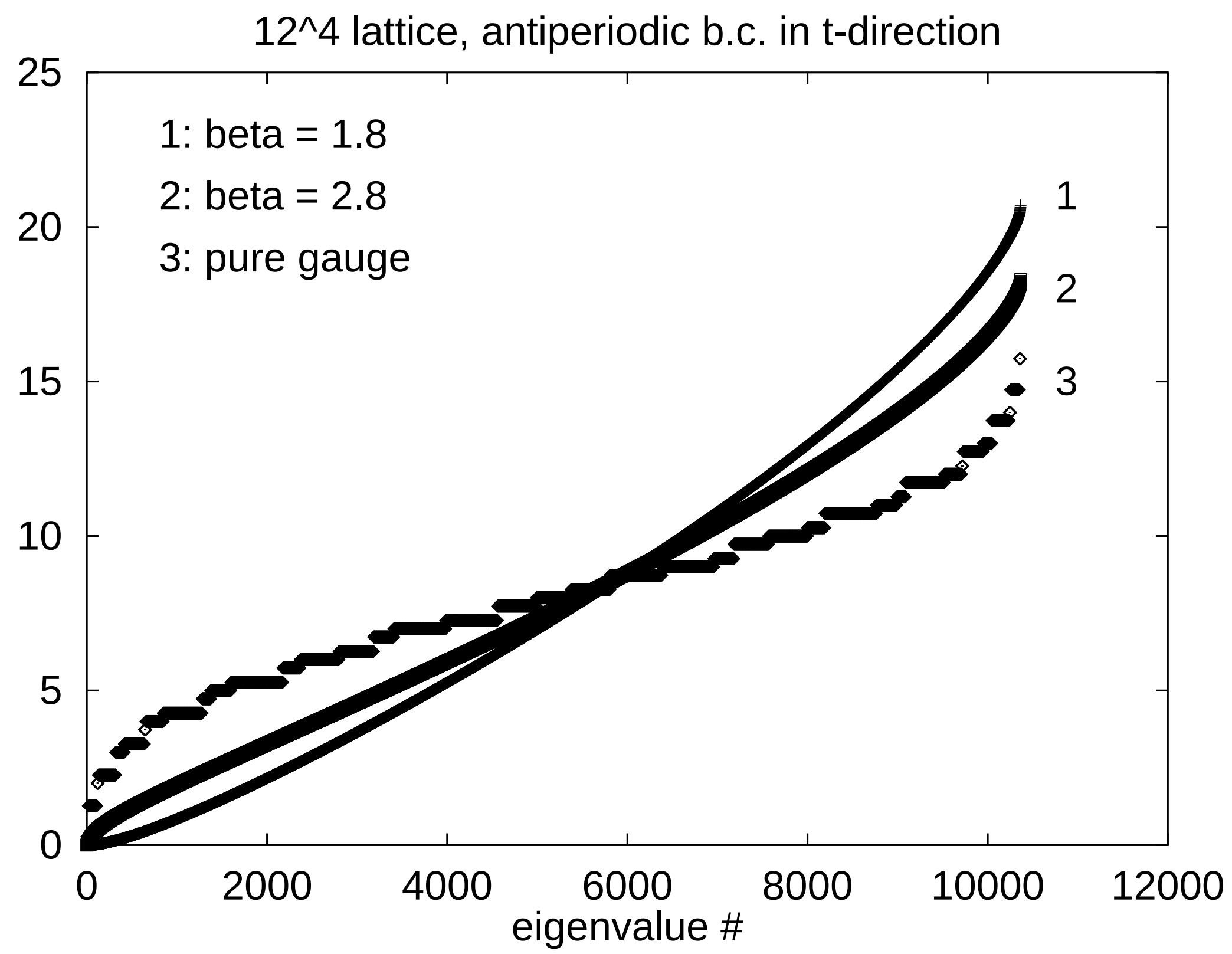

Figure 2: *Spectrum of* $-\not{D}^2$ *on* $12^4$ *lattices with antiperiodic boundary conditions for the Fermi field in t-direction.*

The result on a $(2N)^{d-1} \cdot (2\tau)$ lattice with antiperiodic b.c. in the $\tau$-direction and periodic b.c. in the other directions is

$$\lambda_k = 4\sum_{\mu=1}^{d-1} \sin^2\left(\frac{\pi k_\mu}{N}\right) + 4\sin^2\left(\frac{\pi(2k_\tau + 1)}{2\tau}\right)$$
$$\text{with } k_\mu \in \{0, 1, \ldots, \left[\tfrac{N}{2}\right]\} \text{ for } \mu = 1, \ldots, d-1, \text{ and } k_\tau \in \{0, 1, \ldots, k_{\tau,\max}\}$$
$$\text{where } k_{\tau,\max} = \tau/2 - 1 \text{ if } \tau \text{ is even}, = (\tau - 1)/2 \text{ if } \tau \text{ is odd.} \tag{5}$$

The $\beta = \infty$ values (4) and (5) are shown in Figs. 1 and 2 with their true multiplicities modulo 4 so that they indicate the curve which the numerical data should approach for large $\beta$. When one runs the Lanczos program with free fields, one obtains the eigenvalues (4) and (5) very fast and accurately.

The numerical problem mentioned above which was found for $\beta = 5.0$ is probably due to the fact that the eigenvalues in the large $\beta$ region tend to group in the clusters (4) or (5), respectively, where it is hard to disentangle them numerically.

---

6) $\left[\frac{N}{2}\right] = N/2$ if $N$ is even, and $= (N-1)/2$ if $N$ is odd.

When one looks at the complete spectra, figures for $6^4$ lattices look practically the same on the overall range as Figs. 1 and 2, when one rescales the abscissa by 16. Only for the eigenvalues of the free $-\not{D}^2$ this is not true. These values are collected in Table 1.

| periodic b.c. | | | | | | antiperiodic b.c. | | | | | |
|---|---|---|---|---|---|---|---|---|---|---|---|
| eigenvalue | 0 | 3 | 6 | 9 | 12 | eigenvalue | 1 | 4 | 7 | 10 | 13 |
| degeneracy | 8 | 64 | 192 | 256 | 128 | degeneracy | 16 | 104 | 240 | 224 | 64 |

Table 1: *Spectrum of the free staggered $-\not{D}^2$ on $6^4$ lattices. Degeneracies are given modulo 4.*

### 3.2.2 Low-lying Spectra

Now we take a closer look at the low-lying spectra of $-\not{D}^2$. Due to renormalization effects it is difficult to say for which triples $(\Lambda, \beta, m)$ (+ b.c.) the results can be compared physically. We do not intend to do such a comparison in this exploratory study. Nevertheless, one could possibly discover trends, and it is instructive to look at the results as one keeps two parameters fixed.

Fig. 3 shows the lowest eigenvalues (modulo the fourfold degeneracy mentioned above) of $-\not{D}^2$ on $6^4$ lattices at $\beta = 1.8$. We look at individual configurations, but all Monte Carlo runs were independent. In the quenched case and for $m = 0.2$ one is in the confined chirally broken phase, while for $m = 0.05$ one is just in the deconfined chirally symmetric phase [18]. We note the following. In the confined phase there is little difference in the spectra with periodic and with antiperiodic b.c., while in the deconfined phase (or close to the transition) there is a difference. At this point we also want to add the following observation. The chiral condensate $\langle\overline{\chi}\chi\rangle$ as measured by Meyer's and Pendleton's Monte Carlo program sees no difference between the two kinds of b.c. (within statistical errors), in both phases. In contrast, the Polyakov loop is sensitive to the choice of b.c. Disregarding renormalization effects, Fig. 3 confirms the naive expectation that the effect of dynamical quarks is to rise the low-lying spectrum of $-\not{D}^2$; there are less approximate zero modes the lighter the fermions are.

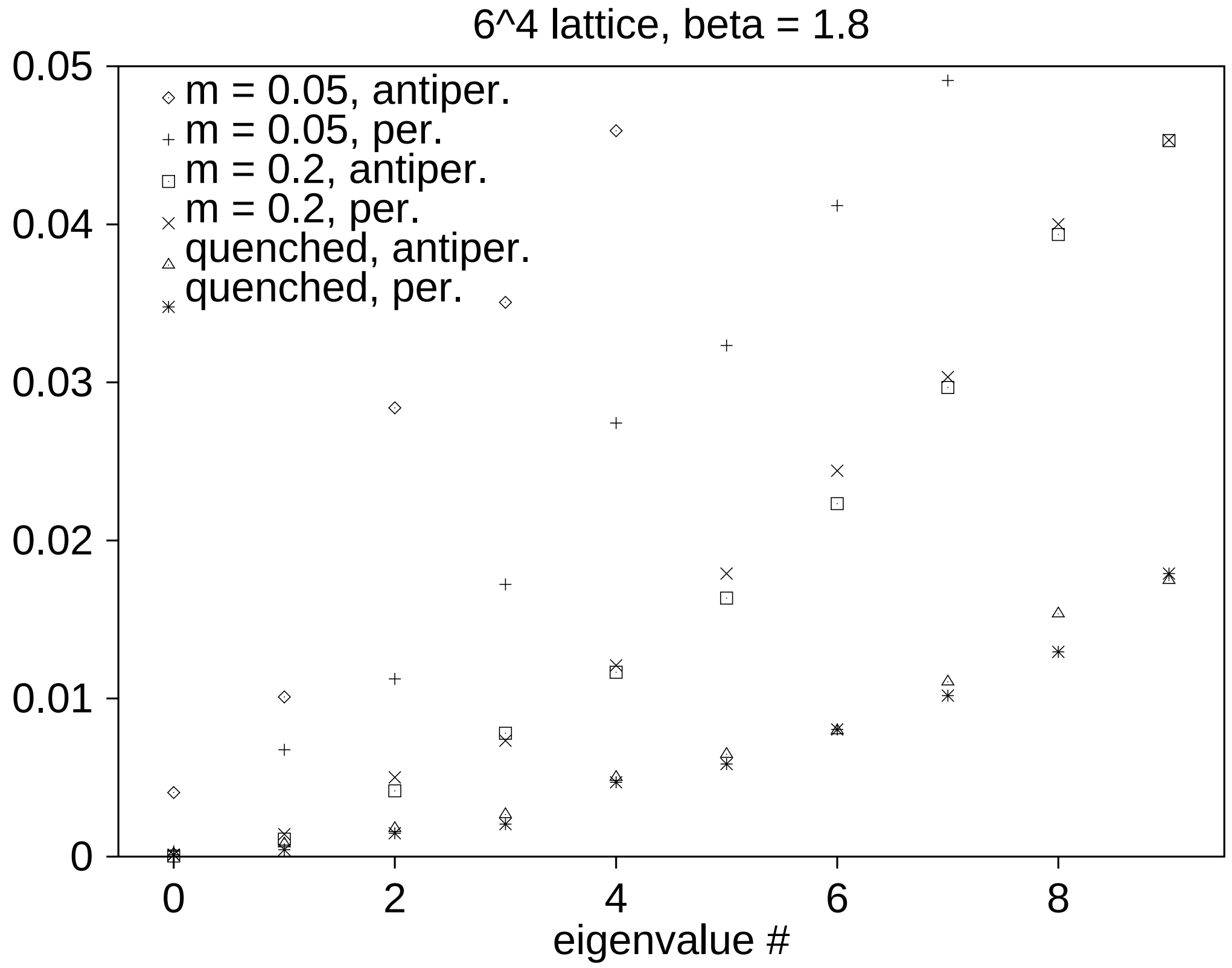

Figure 3: *Low-lying spectrum of $-\not{D}^2$ on $6^4$ lattices at $\beta = 1.8$. $m$ is the quark mass in the Hybrid Monte Carlo program. "antiper." and "per." stand for the choice of boundary conditions. The examples shown are configs. # 1, 2, 5, 6, 9, and 10 of Table 4.*

Figs. 4–6 show examples of the low-lying spectra of the massless operator $-\not{D}^2$ on $12^4$ lattices for the pure gauge theory (with static quarks), and in the presence of dynamical staggered fermions with masses $m = 0.2$ and $m = 0.05$, respectively. Here we note again that the results do not depend on the choice of b.c. for smaller values of $\beta$, while they do for larger $\beta$. In the limiting case of free fields the lowest eigenvalue is always zero in case of periodic b.c. no matter how big the lattice is. In case of antiperiodic b.c. it equals $4\sin^2[\pi/(2\tau)]$, see Table 2; note that in numerical simulations on $6^4$ and $12^4$ lattices the lowest eigenvalue is always much smaller than the free value, it is almost zero.

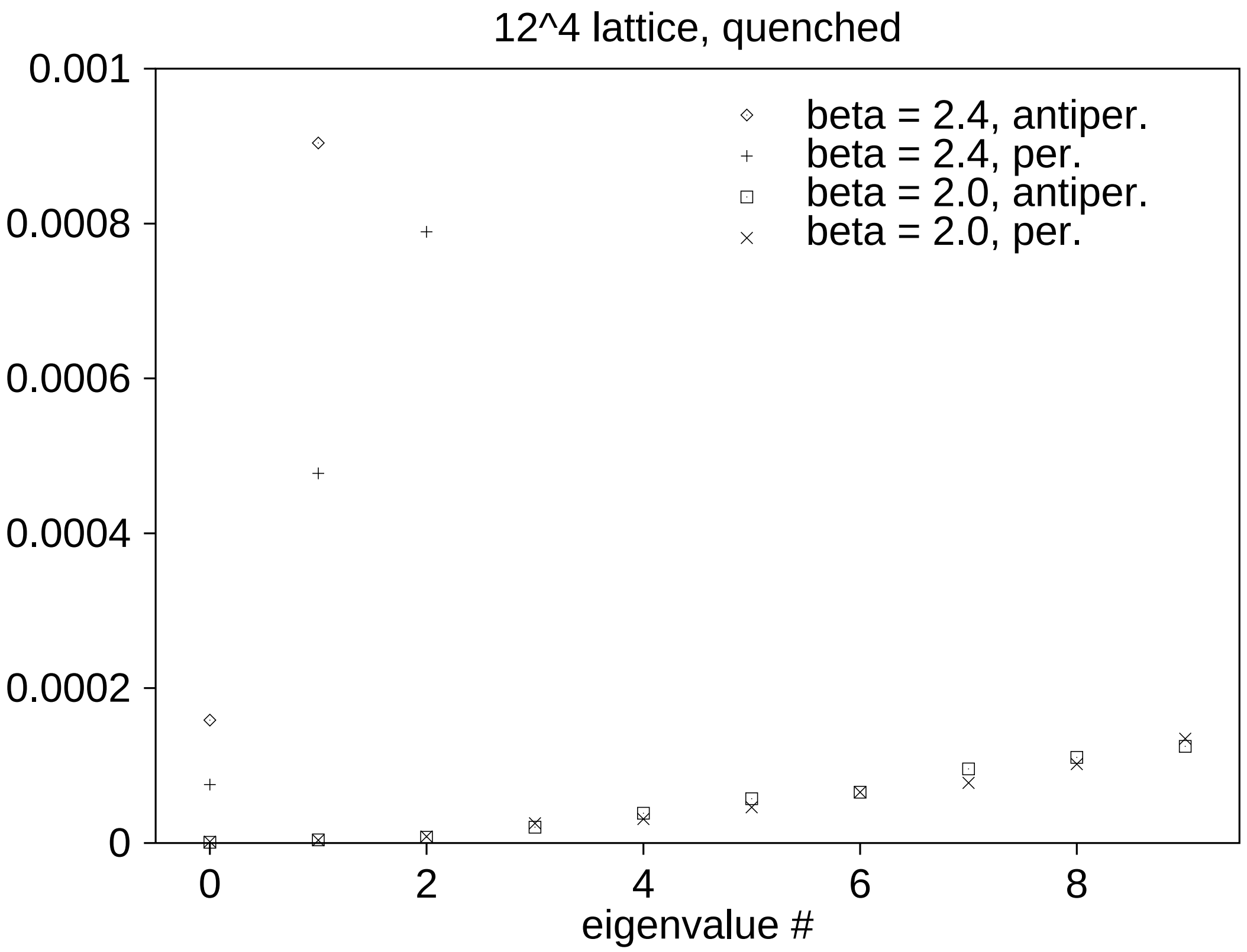

Figure 4: *Low-lying spectra of the quenched operator* $-\not{D}^2$ *on* $12^4$ *lattices. "antiper." and "per." stand for the choice of boundary conditions. Results for* $\beta > 2.4$ *are outside the range of this plot; only for periodic b.c.* $\lambda_0$ *comes back to zero as* $\beta \to \infty$*. The examples shown are configs. # 13–16 of Table 4.*

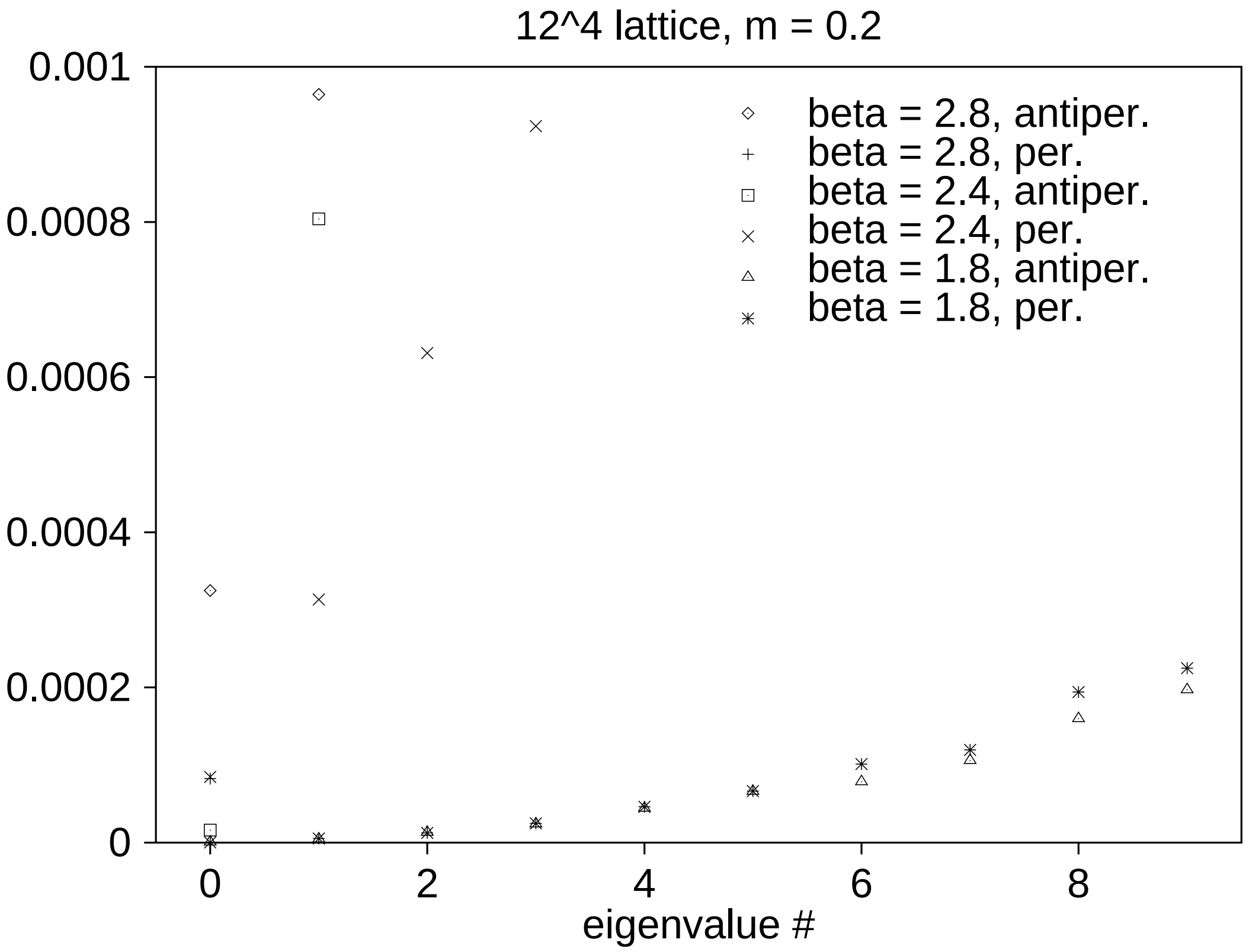

Figure 5: *Low-lying spectra of* $-\not{D}^2$ *on* $12^4$ *lattices in the presence of dynamical staggered fermions of mass* 0.2*. "antiper." and "per." stand for the choice of boundary conditions. (For "*$\beta = 2.8$*, per." only the lowest eigenvalue is visable (i.e. is* $< 0.001$*); this point almost coincides with the one for "*$\beta = 2.4$*, per.".) The examples shown are configs. # 23–28 of Table 4.*

| $\|\Lambda\|$ | $6^4$ | $12^4$ | $18^4$ | $24^4$ | $30^4$ | $36^4$ | $72^4$ |
|---|---|---|---|---|---|---|---|
| $\lambda_0$ | 1.0 | 0.2679491 | 0.1206148 | 0.0681483 | 0.0437048 | 0.0303845 | 0.0076106 |

Table 2: *Lowest eigenvalue of the free* $-\not{D}^2$ *on staggered* $(2L)^{d-1} \cdot (2\tau)$ *lattices with antiperiodic b.c. in* $\tau$*-direction, and periodic b.c. in the* $d-1$ *spatial directions.*

### 3.2.3 Condition numbers

To conclude this section let us give an idea of condition numbers. Call the two masses which were used in the Hybrid Monte Carlo runs $m_1 = 0.1$ and $m_2 = 0.05$. Denote by $\kappa_i$, $i = 1, 2$, the condition number of $(-\not{D}^2 + m_i^2)$, i.e. $\kappa_i = (\lambda_{\max} + m_i^2)/(\lambda_{\min} + m_i^2)$, where $\lambda_{\max}$ and $\lambda_{\min}$ ($\equiv \lambda_0$) denote the highest and the lowest eigenvalue of $-\not{D}^2$, respectively. Results for free fields are in Table 3. Table 4 gives examples in particular nontrivial configurations on $6^4$ and $12^4$ lattices. (Note that it makes no sense to quote $\kappa_2$ for Hybrid Monte Carlo runs with $m_1$, and vice versa.) The configuration numbers ("config. #") are referred to in Sec. 4.

Condition numbers are (much) larger in nontrivial gauge fields than in case of free fields, in particular for antiperiodic b.c. This is just another manifestation that the inversion of $(-\not{D}^2 + m^2)$ in nontrivial gauge fields is much harder than the computation of free propagators. Finally, note that $\lambda_{\max}$ in Table 4 depends in general little on the choice of boundary conditions.

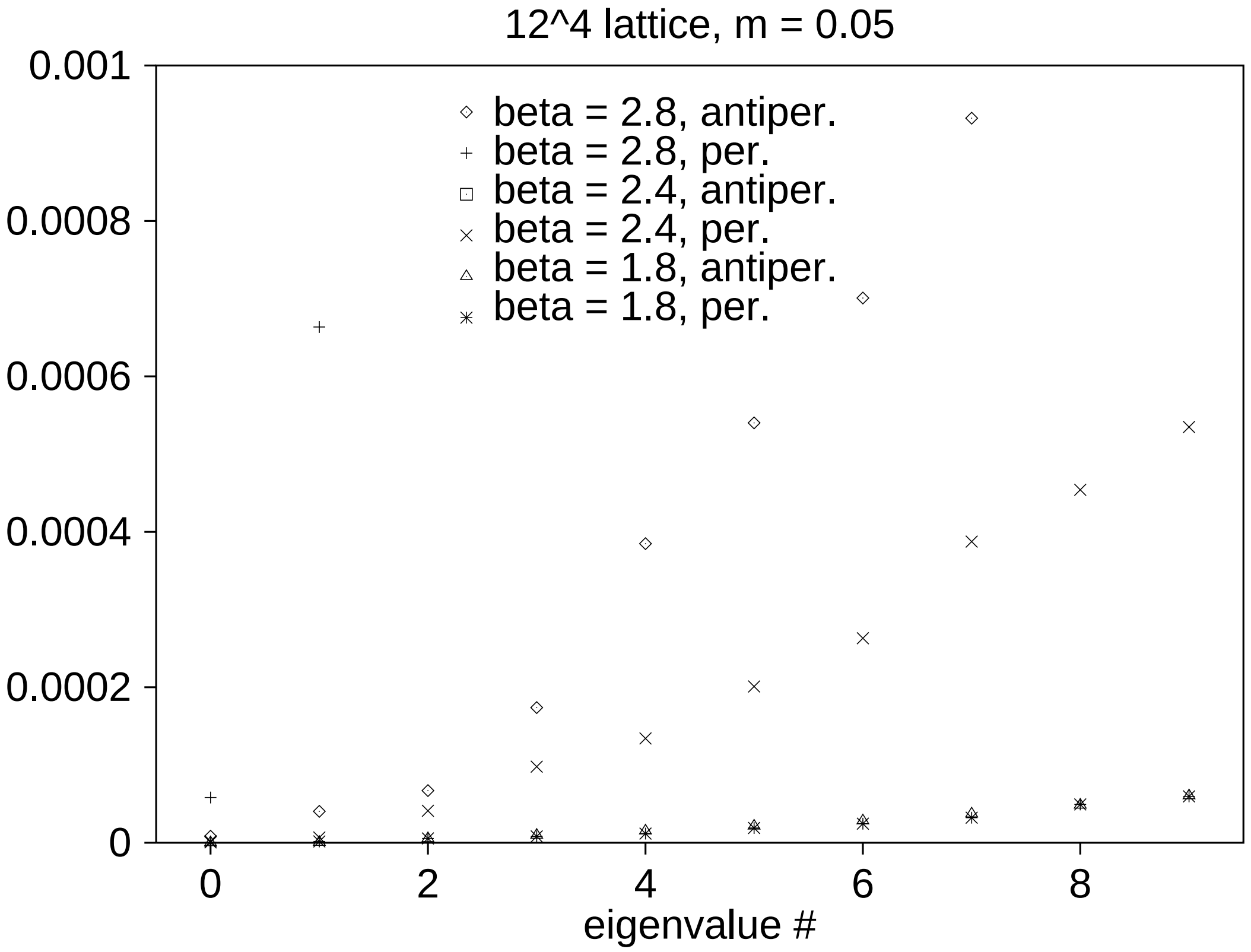

Figure 6: *Low-lying spectra of* $-\not{D}^2$ *on* $12^4$ *lattices in the presence of dynamical staggered fermions of mass* 0.05*. "antiper." and "per." stand for the choice of boundary conditions. (*$\lambda_0 = 0.011$ *for "*$\beta = 2.4$*, antiper.", so that for these parameters no point is visable on the scale of the plot.) The examples shown are configs. # 29–34 of Table 4.*

| $\lvert\Lambda\rvert$ | b.c. | $\lambda_{\rm min}$ | $\lambda_{\rm max}$ | $\kappa_1$ | $\kappa_2$ |
|---|---|---|---|---|---|
| $6^4$ | per. | 0 | 12 | 301 | 4801 |
| $6^4$ | antiper. | 1 | 13 | 12.5 | 13.0 |
| $12^4$ | per. | 0 | 16 | 401 | 6401 |
| $12^4$ | antiper. | 0.26795 | 15.73205 | 51.2 | 58.2 |

Table 3: *Extremal eigenvalues of the free* $-\not{D}^2$*, and condition numbers* $\kappa_i$*.*

| config. # | $\|\Lambda\|$ | b.c. | $m$ | $\beta$ | $\lambda_{\mathrm{min}}$ | $\lambda_{\mathrm{max}}$ | $\kappa_1$ | $\kappa_2$ |
|---|---|---|---|---|---|---|---|---|
| 1 | $6^4$ | per. | $\infty$ | 1.8 | $1.222 \cdot 10^{-4}$ | 21.15 | 518.1 | 8066 |
| 2 | $6^4$ | antiper. | $\infty$ | 1.8 | $1.725 \cdot 10^{-4}$ | 21.14 | 527.2 | 7910 |
| 3 | $6^4$ | per. | $\infty$ | 2.8 | $5.668 \cdot 10^{-2}$ | 18.28 | 189.5 | 309 |
| 4 | $6^4$ | antiper. | $\infty$ | 2.8 | $2.029 \cdot 10^{-1}$ | 18.48 | 76.2 | 90 |
| 5 | $6^4$ | per. | 0.2 | 1.8 | $1.119 \cdot 10^{-4}$ | 20.73 | 517.8 | – |
| 6 | $6^4$ | antiper. | 0.2 | 1.8 | $4.045 \cdot 10^{-5}$ | 20.76 | 519.5 | – |
| 7 | $6^4$ | per. | 0.2 | 2.8 | $4.016 \cdot 10^{-1}$ | 18.25 | 41.4 | – |
| 8 | $6^4$ | antiper. | 0.2 | 2.8 | $4.304 \cdot 10^{-1}$ | 18.38 | 39.2 | – |
| 9 | $6^4$ | per. | 0.05 | 1.8 | $3.395 \cdot 10^{-4}$ | 20.58 | – | 7248 |
| 10 | $6^4$ | antiper. | 0.05 | 1.8 | $4.052 \cdot 10^{-3}$ | 20.36 | – | 3107 |
| 11 | $6^4$ | per. | 0.05 | 2.8 | $3.887 \cdot 10^{-1}$ | 18.38 | – | 47 |
| 12 | $6^4$ | antiper. | 0.05 | 2.8 | $3.990 \cdot 10^{-1}$ | 18.32 | – | 46 |
| 13 | $12^4$ | per. | $\infty$ | 2.0 | $1.747 \cdot 10^{-6}$ | 20.73 | 519.2 | 8286 |
| 14 | $12^4$ | antiper. | $\infty$ | 2.0 | $8.574 \cdot 10^{-7}$ | 20.70 | 518.5 | 8277 |
| 15 | $12^4$ | per. | $\infty$ | 2.4 | $7.556 \cdot 10^{-5}$ | 19.30 | 482.6 | 7494 |
| 16 | $12^4$ | antiper. | $\infty$ | 2.4 | $1.588 \cdot 10^{-4}$ | 19.29 | 481.3 | 7255 |
| 17 | $12^4$ | per. | $\infty$ | 2.6 | $4.204 \cdot 10^{-3}$ | 18.82 | 426.7 | 2807 |
| 18 | $12^4$ | antiper. | $\infty$ | 2.6 | $1.916 \cdot 10^{-2}$ | 18.82 | 318.8 | 869 |
| 19 | $12^4$ | per. | $\infty$ | 2.7 | $4.915 \cdot 10^{-2}$ | 18.66 | 209.8 | 361 |
| 20 | $12^4$ | antiper. | $\infty$ | 2.7 | $2.621 \cdot 10^{-2}$ | 18.66 | 282.4 | 650 |
| 21 | $12^4$ | per. | $\infty$ | 2.8 | $4.002 \cdot 10^{-2}$ | 18.47 | 462.5 | 434 |
| 22 | $12^4$ | antiper. | $\infty$ | 2.8 | $6.198 \cdot 10^{-2}$ | 18.48 | 181.6 | 287 |
| 23 | $12^4$ | per. | 0.2 | 1.8 | $3.075 \cdot 10^{-6}$ | 20.68 | 514.0 | – |
| 24 | $12^4$ | antiper. | 0.2 | 1.8 | $6.911 \cdot 10^{-7}$ | 20.70 | 518.5 | – |
| 25 | $12^4$ | per. | 0.2 | 2.4 | $8.475 \cdot 10^{-5}$ | 18.97 | 474.2 | – |
| 26 | $12^4$ | antiper. | 0.2 | 2.4 | $1.602 \cdot 10^{-5}$ | 19.13 | 479.1 | – |
| 27 | $12^4$ | per. | 0.2 | 2.8 | $8.250 \cdot 10^{-5}$ | 18.31 | 457.8 | – |
| 28 | $12^4$ | antiper. | 0.2 | 2.8 | $3.249 \cdot 10^{-4}$ | 18.30 | 454.8 | – |
| 29 | $12^4$ | per. | 0.05 | 1.8 | $4.222 \cdot 10^{-7}$ | 21.38 | – | 8550 |
| 30 | $12^4$ | antiper. | 0.05 | 1.8 | $9.507 \cdot 10^{-8}$ | 21.55 | – | 8620 |
| 31 | $12^4$ | per. | 0.05 | 2.4 | $1.956 \cdot 10^{-6}$ | 19.73 | – | 7886 |
| 32 | $12^4$ | antiper. | 0.05 | 2.4 | $1.053 \cdot 10^{-2}$ | 18.93 | – | 1453 |
| 33 | $12^4$ | per. | 0.05 | 2.8 | $5.829 \cdot 10^{-5}$ | 18.46 | – | 7216 |
| 34 | $12^4$ | antiper. | 0.05 | 2.8 | $8.323 \cdot 10^{-6}$ | 18.88 | – | 7527 |

Table 4: *Examples for the extremal eigenvalues of $-\not{D}^2$, and for condition numbers $\kappa_i$ in particular gauge fields on $|\Lambda|$ lattices. The value in the column "m" gives the value of the mass of the dynamical fermions in the Hybrid Monte Carlo run; $m = \infty$ stands for a quenched simulation.*

# 4 Inversion of $(-\not{D}^2 + m^2)$

A comprehensive summary about the computation of propagators by means of various algorithms in quenched gauge fields can be found in [8]. Here we focus on the standard CG algorithm [19] and the multigrid method of Sec. 2.

## 4.1 Results of the conjugate gradient algorithm

One often finds the general statement that the speed of convergence of CG depends on the condition number [24]. In cases where the extremal eigenvalues are well separated one can find "superlinear convergence", i.e. convergence at a rate that increases per iteration. More precisely [25], the asymptotic convergence rate of CG depends exclusively on the condition number (i.e. only on the extremal eigenvalues), but the form of the convergence behavior is influenced by the entire spectrum. If the eigenvalues are not distributed uniformly between $\lambda_{\min}$ and $\lambda_{\max}$ (i.e. if they are clustered or there are large degeneracies), then CG converges better than the estimate determined by the condition number.

In case of free fields the eigenvalues of $(-\not{D}^2 + m^2)$ are clustered, Eqs. (4) and (5), and the results of the computation of free propagators by CG [8] may be interpreted as a kind of superlinear convergence. As we saw in Sec. 3, in nontrivial gauge fields the eigenvalues are distributed uniformly between $\lambda_{\min}$ and $\lambda_{\max}$ so that "standard" convergence must be expected. It is already known [8] that the inversion of $(-\not{D}^2 + m^2)$ becomes harder the more disordered the gauge field becomes.

A result of the present study is that the convergence behavior of CG in nontrivial gauge fields is practically only determined by the condition number $\kappa$ of $(-\not{D}^2 + m^2)$, and by the lattice size; see Fig. 7. For configurations on a lattice of given size with the same $\kappa$, CG yields sequences of RMS norms of residuals which practically coincide, even if the spectra are different. This comes as no surprise for configurations where the spectra are almost identical (e.g configs. # 13 and 14, 23 and 24, 29 and 30), but it is also true in cases where there are more differences in the spectra (e.g. configs. # 15 and 16, 25 and 26). However, on the other hand, as mentioned in Sec. 3, on the overall range of the spectra there is little difference between quenched simulations and simulations with dynamical fermions (of mass $m = 0.2, 0.05$). Therefore slight fluctuations in the distrubution of eigenvalues on small scales do not affect the convergence of CG. Thus if one wants to study the convergence of the CG algorithm one can do that with "cheap" quenched gauge fields, one does not have to take "expensive" unquenched configurations.

Finally let us note that identical convergence behavior (measured by RMS norms) was found for $m = 0.2$ on $6^4$ and $12^4$ lattices. Only for the smaller mass $m = 0.05$ the RMS residual is reduced faster on the $6^4$ lattice (Fig. 7).

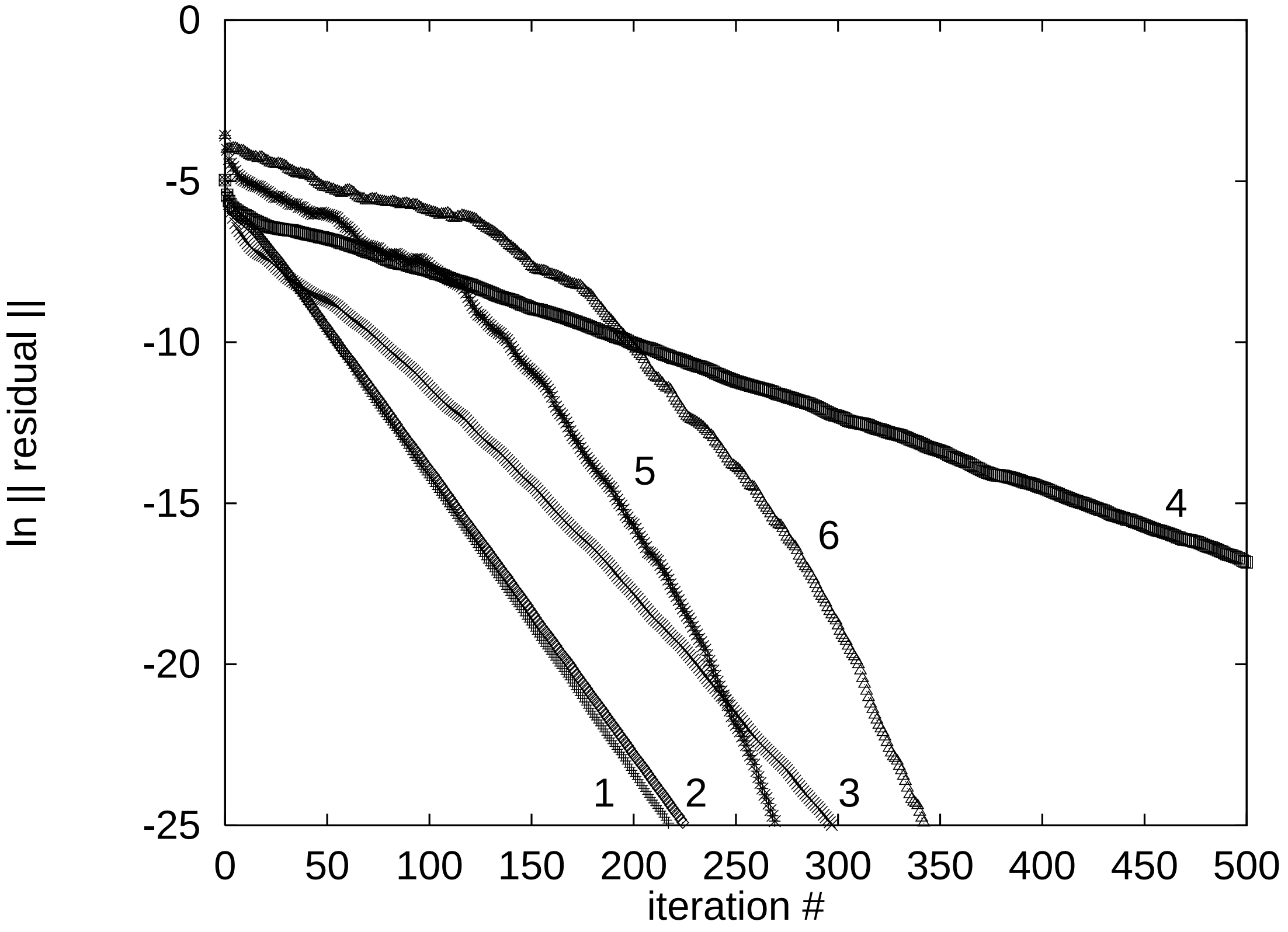

Figure 7: *CG convergence of the RMS residuals in dependence on the condition number* $\kappa$*. Curves 1–4 are results for* $12^4$ *lattices with* $\kappa = 479, 519, 1453, 7886$*, respectively. Curves 5 and 6 are results on* $6^4$ *lattices with* $\kappa = 3107, 7248$*, respectively. The curve for convergence on a* $6^4$ *lattice with* $\kappa = 518$ *coincides with curve # 2.*

## 4.2 Results of the twogrid algorithm

For the inversion of $(-\not{D}^2 + m^2)$ by means of an MG method we used the twogrid algorithm described in Sec. 2, where the relaxation scheme on the fine grid was successive overrelaxation (SOR) with a relaxation parameter $\omega$, and sweeping was done in lexicographic ordering. According to the conventional MG wisdom Gauß-Seidel relaxation ($\omega = 1$) is a good smoother. However, from previous works [8] we know that the picture changes in nontrivial gauge fields. The performance of our simple variational MG method can be improved at finite gauge coupling by choosing $\omega > 1$.

An obvious statement is that convergence of the MG algorithm is not determined by the condition number $\kappa$. This is clear in the limiting case of free fields, because in pure gauges critical slowing down is completely eliminated by MG, i.e. convergence is completely independent of $\kappa$.

In nontrivial gauge fields convergence of MG depends on details of the spectrum. For instance, configs. # 1, 2, 5, and 6 all have the same $\kappa$ for $m = 0.2$ (Table 4). The spectra are practically equal for configs. # 1 and 2, and for configs. # 5 and 6, and so is the MG convergence within each of the two groups (also as a function of $\omega$). But MG convergence in config. # 1 is distinctly different from config. # 5. Convergence was monitored for $\omega = 1.0, 1.2, 1.4, 1.6, 1.8$, and 1.9. In config. #1 the best value was 1.6, while it was 1.8 in config. #5. MG with plain Gauß-Seidel relaxation performed identical on configs. #1 and 5. In all cases evident inferiority of MG was found compared to CG, a factor of about 10 in CPU time.

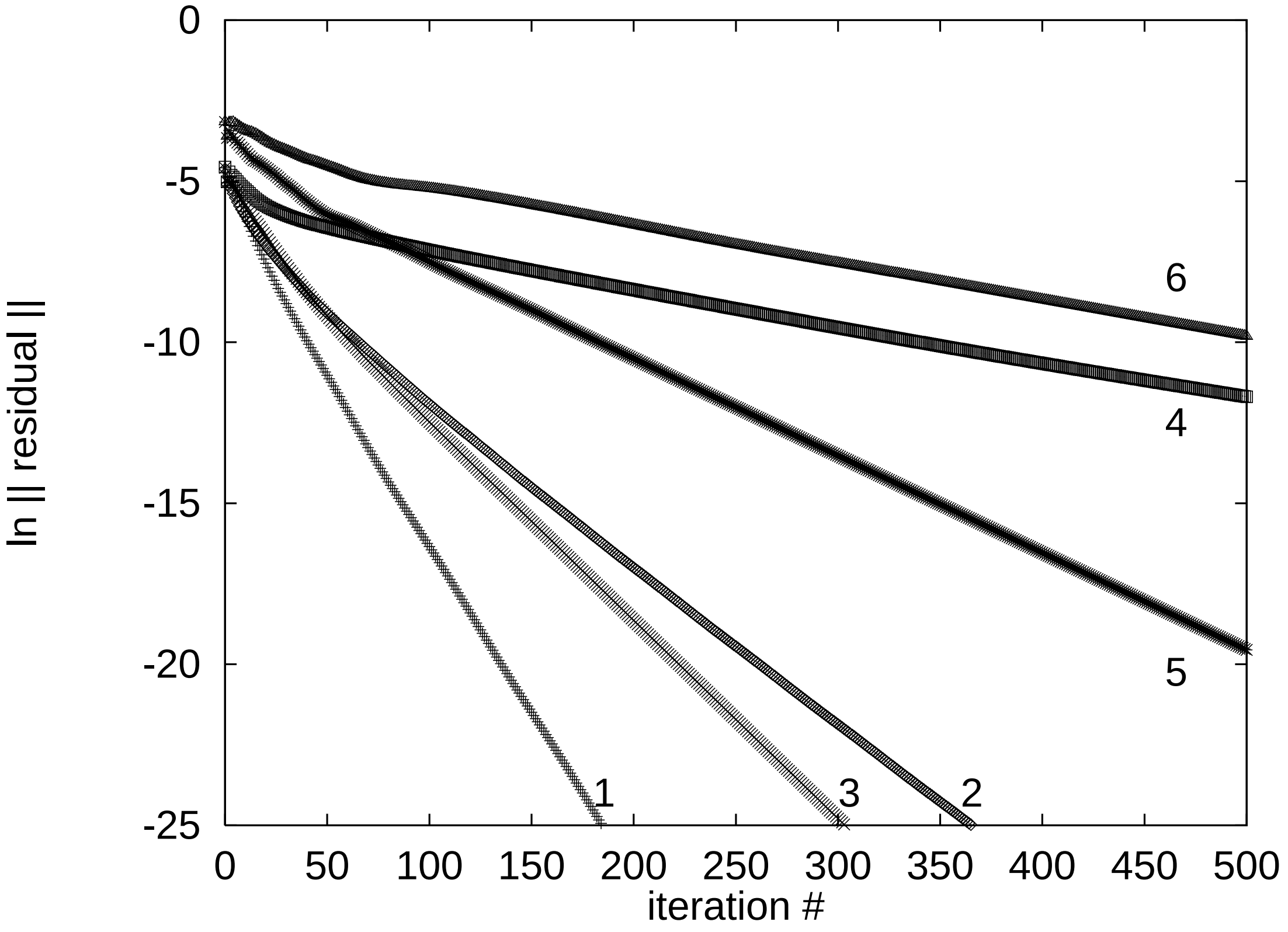

Figure 8: *MG convergence of RMS residuals. The numbers at the curves have the same meaning as in the caption of Fig. 7. The relaxation parameter $\omega$ is* 1.8 *for* $m = 0.2$, *and* 1.9 *for* $m = 0.05$. *Here the curve for convergence on a* $6^4$ *lattice with* $\kappa = 518$ *does not coincide with curve # 2, as it does in Fig. 7.*

On $12^4$ lattices we monitored MG convergence for the same set of $\omega$-values mentioned above, and for all $12^4$ configurations of Table 4. The best $\omega$-value depends on the individual gauge field. Roughly speaking one obtains best convergence if one chooses $\omega = 1.8$ for $m = 0.2$, and $\omega = 1.9$ for $m = 0.05$. But again we could not find any significant difference in the performance of the MG algorithm in quenched and unquenched gauge fields. The poor performance of MG found earlier [8] is no feature of quenched computations.

We conclude by giving results of MG computations in Fig. 8, where convergence is shown for the same configurations as in Fig. 7. We stress that we show convergence in number of iterations. Conversion to CPU time favors CG by another factor of 4.5.

# 5 Conclusions

The complete spectrum of the staggered Dirac operator in four-dimensional $SU(2)$ gauge fields can be determined very accurately by Cullum's and Willoughby's Lanczos procedure [21], provided the Wilson coupling $\beta = 4/g^2$ is not too large. At finite $\beta$ the eigenvalues of $-\not{D}^2$ are distributed uniformly between the lowest and the highest eigenvalue. This is so both for quenched simulations and for simulations with dynamical fermions. On the overall scale the shape of the spectrum depends little on the fermion mass. As a consequence the convergence of the CG algorithm is only determined by the condition number $\kappa$. On a lattice of given size CG produces iterates whose norms depend only on $\kappa$. Since $\kappa$ is almost not affected by the presence of dynamical fermions, one can predict the convergence of CG in unquenched simulations from quenched simulations.

With antiperiodic boundary conditions the lowest eigenvalue of $-\not{D}^2$ is $4\sin^2[\pi/(2\tau)]$ which is not so close to zero on lattices of realistic size. However, when one introduces a nontrivial gauge field the lowest eigenvalue is brought very close to zero. Moreover, for intermediate values of $\beta$ the spectra are practically the same for periodic and for antiperiodic boundary conditions.

On a $6^4$ lattice we found that at fixed $\beta$ the low-lying spectrum is raised when dynamical fermions are introduced, and that this rise is bigger the lighter the mass of the dynamical quarks become. (This is a general trend, also when one passes the finite-temperature phase transition.) Naively this can be taken as a confirmation of the expectation that the effect of dynamical fermions is to suppress configurations with many approximate zero modes. However, one has to consider renormalization effects, which we did not intend to do in this exploratory study. That renormalization effects play an important role can be seen already from the results on $12^4$ lattices.

For the performance of the variational MG method studied here we could not find any improvement when quenched gauge fields are replaced by configurations with dynamical fermions. We could only rediscover the previously known result [8] that there will be a breakeven point in lattice sizes after which MG will outperform CG. This is so because in the limiting case $\beta \to \infty$ critical slowing down is completely eliminated. However, we cannot judge how big the lattices have to be in order to reach the breakeven point. It is reasonable to believe that in principle the presence of dynamical fermions will affect the performance of MG algorithms in a positive way. We think that the main reason for seeing no improvement is that the notion of "Laplace smoothness" [14] which stands behind the definition (3) is inappropriate for staggered fermions. One should rather use the "Dirac notion of smoothness" [14, 10] which is in the spirit of the discussion in Ref. [15, 16]. Possibly with Bäker's algorithm [16] one can observe that the presence of dynamical fermions simplifies the task for MG algorithms.

Acknowledgments

The present work was begun after Achi Brandt pointed out that the inclusion of dynamical fermions might influence MG computations of propagators in a positive way. I wish to thank Achi for discussions in Israel (which were made possible through a travel grant of GIF) and for Email correspondence. I also wish to thank Gerhard Mack and Steffen Meyer for many stimulating discussions. Special thanks are due to Steffen Meyer for providing me with the Hybrid Monte Carlo program which he used in joint work with Brian Pendleton. Tony Kennedy is thanked for an Email correspondence on trajectory lengths in Hybrid Monte Carlo simulations. Financial support by Deutsche Forschungsgemeinschaft is gratefully acknowledged. The computations reported here were performed on the CRAY Y-MP of the University of Kaiserslautern.

# References

[1] R. Ben-Av, M. Harmatz, P.G. Lauwers, and S. Solomon, *Parallel-transported multigrid for inverting the Dirac operator: variants of the method and their efficiency*, Nucl. Phys. B405 (1993) 623, and references therein;

P.G. Lauwers and T. Wittlich, *Inversion of the fermion matrix in lattice QCD by means of parallel-transported multigrid (PTMG)*, Int. J. Mod. Phys. C4 (1993) 609.

[2] M. Harmatz, P. Lauwers, S. Solomon, and T. Wittlich, *Visual study of zero-modes role in PTMG convergence*, Nucl. Phys. B (Proc. Suppl.) 30 (1993) 192.

[3] R.C. Brower, R.G. Edwards, C. Rebbi, and E. Vicari, *Projective multigrid for Wilson fermions*, Nucl. Phys. B366 (1991) 689, and references therein.

[4] A. Hulsebos, J. Smit, and J.C. Vink, *Multigrid inversion of lattice fermion operators*, Nucl. Phys. B368 (1992) 379, and references therein;

J.C. Vink, *Multigrid inversion of fermion operators with* $SU(2)$ *gauge fields in two and four dimensions*, Nucl. Phys. B (Proc. Suppl.) 26 (1992) 607.

[5] V.Vyas, *A multigrid algorithm for calculating fermionic propagators using Migdal-Kadanoff transformation*, Wuppertal preprint WUB 91-10 (February 1991);

*An efficient algorithm for calculating the quark propagators using Migdal-Kadanoff transformation*, Wuppertal preprint WUB 92-30 (September 1992).

[6] T. Kalkreuter, *Ground-state projection multigrid for propagators in four-dimensional* $SU(2)$ *gauge fields*, Phys. Lett. B276 (1992) 485.

[7] T.Kalkreuter, *Multigrid for propagators of staggered fermions in four-dimensional* $SU(2)$ *gauge fields*, Nucl. Phys. B (Proc. Suppl.) 30 (1993) 257;

*Improving multigrid and conventional relaxation algorithms for propagators*, Int. J. Mod. Phys. C3 (1993) 1323.

[8] T. Kalkreuter, *Multigrid Methods for the Computation of Propagators in Gauge Fields*, Ph. D. thesis and preprint DESY 92–158 (November 1992), shortened version to appear in Int. J. Mod. Phys. C5.

[9] T. Kalkreuter, *Idealized multigrid algorithm for staggered fermions*, Phys. Rev. D48 (1993) 1926.

[10] T. Kalkreuter, *Towards multigrid methods for propagators of staggered fermions with improved averaging and interpolation operators*, Nucl. Phys. B (Proc. Suppl.) 34 (1994) 768.

[11] S. Duane, A.D. Kennedy, B.J. Pendleton, and D. Roweth, *Hybrid Monte Carlo*, Phys. Lett. B195 (1987) 216.

[12] M. Lüscher, *A new approach to the problem of dynamical quarks in numerical simulations of lattice QCD*, Nucl. Phys. B418 (1994) 637.

[13] H. Joos and M. Schaefer, *The representation theory of the symmetry group of lattice fermions as a basis for kinematics in lattice QCD*, Z. Phys. C34 (1987) 465.

[14] T. Kalkreuter, G. Mack, and M. Speh, *Blockspin and multigrid for staggered fermions in non-abelian gauge fields*, Int. J. Mod. Phys. C3 (1992) 121.

[15] M. Bäker, T. Kalkreuter, G. Mack, and M. Speh, *Neural multigrid methods for gauge theories and other disordered systems*, in: Proceedings of the 4th International Conference on Physics Computing PC '92, eds. R.A. de Groot and J. Nadrchal (World Scientific, Singapore, 1993);

M. Bäker, G. Mack, and M. Speh, *Multigrid meets neural nets*, Nucl. Phys. B (Proc. Suppl.) 30 (1993) 269.

[16] M. Bäker, *Localization in lattice gauge theory and a new multigrid method*, preprint DESY 94-079 (May 1994).

[17] T. Kalkreuter, *Projective block spin transformations in lattice gauge theories*, Nucl. Phys. B376 (1992) 637.

[18] S. Meyer and B. Pendleton, *First order phase transition in $SU(2)$ gauge theories with many flavours of staggered fermions*, Phys. Lett. B241 (1990) 397;

*The chiral transition in the $SU(2)$ Higgs model at weak coupling*, Phys. Lett. B253 (1991) 205.

[19] G.H. Golub and C.F. v. Loan, *Matrix computations*, second edition, (The Johns Hopkins University Press, Baltimore, 1990).

[20] I.M. Barbour, N.-E. Behilil, P.E. Gibbs, G. Schierholz, and M. Teper, *The Lanczos method in lattice gauge theories*, in: The recursion method and its applications, Springer Series in Solid-State Sciences 58, eds. D. G. Pettifor and D. L. Weaire (Springer-Verlag, Berlin, 1985).

[21] J. Cullum and R.A. Willoughby, *Computing eigenvalues of very large symmetric matrices – an implementation of a Lanczos algorithm with no reorthogonalization*, J. Comp. Phys. 44 (1981) 329.

[22] NAGLIB, Numerical Algorithms Group, NAG Fortran (Vector) Library, Mark 15.

[23] J. Stoer, *Einführung in die Numerische Mathematik I* (Springer-Verlag, Berlin, 1979).

[24] R. Barrett et al., *Templates for the solution of linear systems: building blocks for iterative methods*, obtainable via anonymous ftp from `netlib2.cs.utk.edu`, file `linalg/templates.ps`.

[25] W. Hackbusch, *Iterative Lösung großer schwachbesetzter Gleichungssysteme* (B.G. Teubner, Stuttgart, 1991).